\magnification\magstep1




\overfullrule=0pt  
\hbadness=10000      
\vbadness=10000

\font\eightit=cmti8  
\font\eightrm=cmr8 \font\eighti=cmmi8                 
\font\eightsy=cmsy8 
\font\sixrm=cmr6


\def\eightpoint{\normalbaselineskip=10pt 
\def\rm{\eightrm\fam0} \let\it\eightit
\textfont0=\eightrm \scriptfont0=\sixrm 
\textfont1=\eighti \scriptfont1=\seveni
\textfont2=\eightsy \scriptfont2=\sevensy 
\normalbaselines \eightrm
\parindent=1em}



\def\eq#1{{\noexpand\rm(#1)}}          
\newcount\eqcounter                    
\eqcounter=0                           
\def\numeq{\global\advance\eqcounter by 1\eq{\the\eqcounter}}           
\def\relativeq#1{{\advance\eqcounter by #1\eq{\the\eqcounter}}}



\def\namelasteq#1{\global\edef#1{{\eq{\the\eqcounter}}}}  


\def\A{{\rm A}}
\def\Aslash{{ {\rm A}\mkern-11mu/}}



\def\cite#1{{\rm[#1]}}                 

\def\D{{\rm D}}

\def\Dslash{{ {\rm D}\mkern-11mu/}}    
\def\F{{\rm F}\,}

\def\gam5{\gamma_5}                    






\def\intd{\int\! {\rm d}}              
\def\j{{\rm j}}



\def\nocorr{\kern0pt}                  

\def\pr{\partial}                      
                     

\def\psipsiaction{\bar\psi\,
		\Dslash\,\psi}

       
\def\pslash{{p\mkern-8mu/}{\!}}        
\def\prslash{{\partial\mkern-9mu/}}    





\def\cite#1{{\rm[#1]}}                 



\newif\ifstartsec                   

\outer\def\section#1{\vskip 0pt plus .15\vsize \penalty -250
\vskip 0pt plus -.15\vsize \bigskip \startsectrue
\message{#1}\centerline{\bf#1}\nobreak\noindent}

\def\subsection#1{\ifstartsec\medskip\else\bigskip\fi \startsecfalse
\noindent{\it#1}\penalty100\medskip}

\def\refno#1. #2\par{\smallskip\item{\rm\lbrack#1\rbrack}#2\par}


\def\ABJ{1}
\def\Marshak{2}
\def\Booss{3}
\def\Filk{4}
\def\Minwallaetal{5}
\def\miscellanea{6}
\def\stringt{7}
\def\SW{8}
\def\ArdalanS{9}
\def\Fujikawa{10}
\def\Manolo{11}
\def\Phobos{12}
\def\LAGPG{13}
\def\Cordelia{14}
\def\WZ{15}
\def\Sheikh{16}
\def\Hayakawa{17}


\rightline{FT/UCM--20--2000}

\vskip 1cm

\centerline{\bf Chiral gauge anomalies on Noncommutative ${\rm{I\!R}}^4$}

\bigskip

\centerline{\rm J. M. Gracia-Bond\'{\i}a${}^*$ and C. P.
Mart{\'\i}n${}^{\dag}$}
\medskip
\centerline{\eightit Departamento de F{\'\i}sica Te\'orica I,				
		Universidad Complutense, 28040 Madrid, Spain}
\vfootnote*{email: {\tt jmgb@elbereth.fis.ucm.es}}
\vfootnote\dag{email: {\tt carmelo@elbereth.fis.ucm.es}}

\bigskip\bigskip

\begingroup\narrower\narrower
\eightpoint
We discuss the noncommutative counterparts of chiral gauge theories and
compute the associated anomalies. 
\par
\endgroup 
\vskip 2cm

\section{1. Introduction}

The chiral anomalies~\cite{\ABJ} of quantum field theory are of the
utmost importance in particle physics~\cite{\Marshak} and 
mathematics~\cite{\Booss}.
In this paper we reappraise them in the framework of quantum field
theory on noncommutative manifolds, the latest new hunting ground in both
field theory ---see~\cite{\Filk,\Minwallaetal,\miscellanea}
and string theory ---see~\cite{\stringt,\SW} and the other references 
in this last paper. We shall be concerned mainly
with fermions on noncommutative ${\rm{I\!R}}^4$, coupled to background
$U(N)$ gauge fields. It is good training to start by considering the
axial anomaly; there some of our results overlap with the recent
calculations by Ardalan and Sadooghi~\cite{\ArdalanS}. Then we proceed
to compute the ``dangerous''  anomaly associated to chiral fermions in
nonabelian gauge fields. Variants of the Fujikawa
method~\cite{\Fujikawa,\Manolo}
will be found to be ideally suited for both kinds of computations.

\section{2. The ABJ anomaly on noncommutative ${\rm {I\!R}}^{2n}$}

Noncommutative ${\rm {I\!R}}^{2n}$ is characterized by the Moyal product
of functions
$$
f \star g(x) :=
\int_{\rm {I\!R}^{2n}}\int_{\rm {I\!R}^{2n}}f(x+s)g(x+t)e^{is\omega
t}\,ds\,dt,
$$
where $\omega$ denotes a real antisymmetric matrix, that in this
paper will be taken constant. In general we omit reference to
$\omega$ in the notation. Several straightforward mathematical properties
of that operation will be used in the sequel without further mention, in
particular the Leibniz rule
$$  
\partial_\mu(f \star g) = \partial_\mu f \star g + f \star \partial_\mu
g,
$$
that follows from derivation under the integral sign, and the fact that
$$
\int_{\rm {I\!R}^{2n}} f \star g(x)  \,dx = \int_{\rm {I\!R}^{2n}}
f(x)g(x) \,dx,
$$
inducing cyclicity in the integral of the Moyal product of several
functions. Note also that if $f,g$ decay quickly at infinity, this is
the case for $f \star g$ as well~\cite{\Phobos}.

Consider a $U(1)$ gauge field $A_{\mu}$
on noncommutative ${\rm {I\!R}}^{2n}$. Its infinitesimal gauge variation
is given by $\delta_{\theta}A_{\mu}(x) = \partial_{\mu}\theta(x) -
i[A_{\mu},\theta](x)$, with $\theta(x)$ denoting the gauge parameter and
$[.,.]$ the Moyal commutator or bracket. Let $\psi$ denote an Euclidean
Dirac fermion field on noncommutative ${\rm {I\!R}}^{2n}$, that we take
massless. The fact that the following identity should hold for   
any two $U(1)$ gauge transformations of $\psi$
$$
(\delta_{\theta_1}\delta_{\theta_2}-\delta_{\theta_2}\delta_{\theta_1})\psi=-i
\delta_{[\theta_1,\theta_2]}\psi,
$$
leads {\it uniquely\/} to the following permissible types of gauge
transformations for $\psi$:
$$
a)\;
\delta_{\theta}\psi=i\theta\star\psi,\quad 
b)\;\delta_{\theta}\psi=-i\psi\star\theta,\quad
c)\;\delta_{\theta}\psi=-i[\psi,\theta].
$$
The previous transformations of the fermion field give rise, respectively,
to the
following covariant derivatives
$$
a)\;{\rm D}_{\mu} \psi = \partial_\mu\psi - i A_\mu\star\psi,\quad 
b)\;{\rm D}_{\mu} \psi = \partial_\mu\psi + i \psi\star A_\mu,\quad
c)\;{\rm D}_{\mu} \psi = \partial_\mu\psi - i [A_\mu,\psi].
$$
These yield three types of Dirac $-i\Dslash$ operators twisted
with the $U(1)$ connection on noncommutative ${\rm {I\!R}}^{2n}$, namely
the given by
$$
a)\;\Dslash\, \psi = \prslash\psi - i \gamma_{\mu}A^\mu\star\psi,\; 
b)\;\Dslash\, \psi = \prslash\psi + i \gamma_{\mu}\psi\star A^\mu,\;
c)\;\Dslash\, \psi = \prslash\psi - i \gamma_{\mu}[A^\mu,\psi].
\eqno\numeq\namelasteq\UoneDiracoperators
$$
The latter we exclude from our considerations, as we interested in the
case that the local gauge action occurs in the fundamental
representation of the fermions; in the formula the symbols $\gamma_{\mu}$
denote of course the Euclidean gamma matrices, which satisfy
$\{\gamma_{\mu},\gamma_{\nu}\} = \delta_{\mu\nu}$ and
$\gamma_{\mu}^{\dag} = \gamma_{\mu}$, and the slash notation has its
usual meaning.

Each Dirac operator yields a fermionic action:
$$
{\rm S}_{U(1)}\;=\;-i\intd^{2n}x\;\psipsiaction.
\eqno\numeq\namelasteq\Uoneaction
$$ 
To apply Fujikawa's method to the computation of the  ABJ anomaly for
the quantum theory defined by the previous fermionic action, one begins
by introducing chiral local transformations for the Dirac operators
introduced in equation~\UoneDiracoperators. These transformations
read 
$$
a)\;\delta^5_{\theta}\psi=i\theta\star\gamma_{5}\psi,\quad 
b)\;\delta^5_{\theta}\psi=-i\gamma_{5}\psi\star\theta,\quad
\eqno\numeq\namelasteq\Uonechiraltrans
$$
respectively. The symbol $\gamma_{5}$ denotes the product 
$(-i)^n \prod_{i=1}^{2n} \gamma_{\mu}$. 

Let us consider first the theory
defined by case $a)$ of~\UoneDiracoperators . Then, the variation of
the action in formula~\Uoneaction\ under the chiral transformations $a)$
of~\Uonechiraltrans\ reads
$$
\delta^5_{\theta}{\rm S}_{U(1)}\;=\;\intd^{2n}x\;\theta\big(
\D_{\mu}(A)\j^5_{\mu}\big),
$$ 
where
$$
\j^5_{\mu}\,:=\,\psi_{\beta}\star\bar{\psi}_{\alpha}
(\gamma_{\mu}\gamma_5)_{\alpha\beta},
\quad\D_{\mu}(A)\j^5_{\mu}=\pr_{\mu}\j^5_{\mu}-i[A_{\mu},\j^5_{\mu}].
\eqno\numeq\namelasteq\chiralcurrent
$$
To compute the change of the path integral fermionic measure under the
chiral transformations $a)$ of equation~\Uonechiraltrans, one expands the 
fermionic fields $\psi$ and $\bar{\psi}$ in an orthonormal base of 
eigenfunctions, $\{\varphi_n\}$, of the Dirac operator defined in $a)$ 
of equation~\UoneDiracoperators: 
$$
\psi(x)=\sum_{n}\,a_n\,\varphi_n(x),\quad 
\bar{\psi}=\sum_{n}\,\bar{b}_n\,\varphi^{\dag}_n(x).
$$
The coefficients $a_n$ and $b_n$ are anticommuting variables.  Now, the
standard calculation of the functional derivative of the determinant of
the change of variables in the path integral expression of the partition
function leads to 
$$
\big<\D_{\mu}(A)\j^5_{\mu}(x)\big> \,=\, i{\cal A}(x),\quad {\rm
where}\quad {\cal A}(x) =  2\sum_{m}\gamma_{5\,\alpha\beta}
(\varphi_{n\beta}\star\varphi^{\dag}_{n\alpha})(x).
\eqno\numeq\namelasteq\Nonconserved
$$
The sum ${\cal A}(x)$ is not well-defined and to
define it we first regularize it in a gauge invariant way with the help 
of the heat kernel associated with the Dirac operator at hand. Hence,
$$
{\cal A}(x)\,=\,\lim_{\Lambda\rightarrow\infty}\;
 2\sum_{m}\gamma_{5\,\alpha\beta}\bigg(
\big(e^{\Dslash^2\over \Lambda^2}\varphi_{n}\big)_{\beta}
\star\varphi^{\dag}_{n\alpha}\bigg)(x).
$$
By expanding $\varphi_{n}$ in plane waves, one can recast ${\cal A}(x)$
into
the form
$$
{\cal A}(x)\,=\,\lim_{\Lambda\rightarrow\infty}\;2\int 
{d^{2n}p \over {(2\pi)}^{2n}}
{\rm tr}\bigg(\gamma_{5}
\big(e^{\Dslash^2\over \Lambda^2}e^{ipx}\big)
\star e^{-ipx}\bigg).
$$
A little computation yields
$$
{\cal A}(x)=\lim_{\Lambda\rightarrow\infty}\;2\int\!\! {d^{2n}p \over
{(2\pi)}
^{2n}}{\rm tr}\bigg(\gamma_{5}
\sum_{m}
\bigg({\bigl[(ip_\mu+\partial_\mu-iA_\mu\star)^2-{i\over 2}
\sigma_{\mu\nu}\F_{\mu\nu}\star\bigr]^{m}\over m!\,\Lambda^{2m}}
{{\rm I}\!\!{\rm I}}\bigg)\star e^{ipx}\star e^{-ipx}\bigg),
$$
where ${\rm tr}$ denotes trace over the Dirac matrices,
$\F_{\mu\nu}=\partial_{\mu}A_{\mu}-\partial_{\nu}A_{\mu}-i
[A_{\mu},A_{\nu}]$,
$\sigma_{\mu\nu}={1\over 2}[\gamma_\mu,\gamma_\nu]$ and
${{\rm I}\!\!{\rm I}}$ denotes the unit function on ${\rm {I\!R}}^{2n}$.

Since $e^{ipx}\star e^{-ipx}={{\rm I}\!\!{\rm I}}$, one finally obtains
that
$$
{\cal A}(x)=2{1\over (4\pi)^n}{1\over n!}
\varepsilon^{\mu_1\cdots\mu_{2n}}
\;\F_{\mu_1\mu_2}\star \F_{\mu_3\mu_4}\star\cdots \star 
\F_{\mu_{2n-1}\mu_{2n}},
\eqno\numeq\namelasteq\UoneABJanomaly
$$
which is a $\star$-deformation of the standard result for commutative 
${\rm {I\!R}}^{2n}$. Substituting the previous result back in 
equations~\Nonconserved, we conclude that the $U(1)$ axial current  
$\j^5_{\mu}(x)$, which is classically (covariantly) conserved, is 
not conserved at the quantum level, i.e., there is an anomaly.

The equation for the ABJ anomaly for the Dirac operator $b)$  of
formula~\UoneDiracoperators\ can be obtained from the previous
expressions by doing first the replacements
$\star_{\omega}\rightarrow\star_{-\omega}$, 
$A_{\mu}\rightarrow -A_{\mu}$ and then undoing them. One then obtains
$$ 
\big<\D_{\mu}(\A)j^5_{\mu}\big> = -2i{(-1)^n\over (4\pi)^n}{1\over n!} 
\varepsilon^{\mu_1\cdots\mu_{2n}}
\;
\F_{\mu_1\mu_2}\star \F_{\mu_3\mu_4}\star\cdots \star
\F_{\mu_{2n-1}\mu_{2n}},
\eqno\numeq\namelasteq\UoneABJanomalyb
$$
where now
$$
\j^5_{\mu}=\bar{\psi}_{\alpha}\star\psi_{\beta}
(\gamma_{\mu}\gamma_5)_{\alpha\beta}
\quad{\rm and} \quad
\D_{\mu}(\A)\j^5_{\mu}=\pr_{\mu}j^5_{\mu}+i[\j^5_{\mu},A_\mu].
\eqno\numeq\namelasteq\theothercurrent
$$
The field strength tensor $\F_{\mu\nu}$ keeps its previous definition.
Notice that the chiral currents defined in equations~\chiralcurrent\
and~\theothercurrent\ differ only by a sign in the commutative limit
(recall the Grassmann character of the classical spinor field); not so
on noncommutative Euclidean space.

Now, it will not come unexpectedly that if the gauge group $U(1)$ is
replaced by
$U(N)$, the singlet currents corresponding to equations~\chiralcurrent\
and~\theothercurrent\ satisfy, respectively, the following anomalous
equations:
$$
\eqalignno{
a) \qquad&\big<\pr_{\mu}\j^5_{\mu}-\sum_{i,j}( \A_{\mu\,ij}\star
\j^5_{\mu\,ji} - \j^5_{\mu\,ij}\star \A_{\mu\,ji}) \big> = \cr 
&2{(-i)^{n-1}\over (4\pi)^n}{1\over n!} \varepsilon^{\mu_1\cdots\mu_{2n}}
\;
{\rm Tr}\,\F_{\mu_1\mu_2}\star \F_{\mu_3\mu_4}\star\cdots \star 
\F_{\mu_{2n-1}\mu_{2n}},\cr
b) \qquad&\big< \pr_{\mu}\j^5_{\mu}+\sum_{i,j}(\A_{\mu\,ij}\star
\j^5_{\mu\,ji}-
\j^5_{\mu\,ij}\star \A_{\mu\,ji}) \big> = \cr &2{i^{n-1}\over
(4\pi)^n}{1\over n!} \varepsilon^{\mu_1\cdots\mu_{2n}}
\;
{\rm Tr}\,\F_{\mu_1\mu_2}\star \F_{\mu_3\mu_4}\star\cdots \star 
\F_{\mu_{2n-1}\mu_{2n}},&\numeq\cr
}\namelasteq{\singletanomaly}
$$
where $\A_{\mu}$ is the U(N) gauge field (taken to be antihermitian) 
$i,j$ are U(N) indices, Tr is the trace on U(N) and $\j^5_{\mu}$ and
$\j^5_{\mu\,ij}$ in $a)$ and $b)$ are given by
$$
\eqalignno{a)\quad &\j^5_{\mu}=\sum_{i}\psi_{\beta\,i}
\star\bar{\psi}_{\alpha\,i}
(\gamma_{\mu}\gamma_5)_{\alpha\beta},\quad\
\j^5_{\mu\,ij}=\psi_{\beta\,i}\star\bar{\psi}_{\alpha\,j}
(\gamma_{\mu}\gamma_5)_{\alpha\beta},\cr
b)\quad&\j^5_{\mu}=\sum_{i}\,\bar{\psi}_{\alpha i}\star\psi_{\beta i}
(\gamma_{\mu}\gamma_5)_{\alpha\beta},\quad 
\j^5_{\mu\, ij}=\bar{\psi}_{\alpha\,i}\star\psi_{\beta\,j}
(\gamma_{\mu}\gamma_5)_{\alpha\beta},&\numeq\cr
}\namelasteq{\singletcurrents}
$$
respectively. Note that equations~\UoneABJanomaly ,\UoneABJanomalyb\ and
\singletanomaly\ would result from a ``na\"{\i}ve" $\star$-deformation of
their counterparts on commutative Euclidean space. Again, cases $a)$
and $b)$ essentially match in the commutative limit.
 
We will now turn our attention to the index of the Dirac
operators $a)$ and $b)$ of equation~\UoneDiracoperators\ with a
$iA_{\mu}$ replaced with a $U(N)$ gauge field $\A_{\mu}$. We can readily
adapt the computational method displayed above to carry out a
physicist's computation of the index. The results read
$$
\eqalign{a)\quad&{\rm index}\;(-i\Dslash)\,=\,
{i^n\over (4\pi)^n}{1\over n!}\;\int d^{2n}x\, 
\varepsilon^{\mu_1\cdots\mu_{2n}}
\;
{\rm Tr}\,\F_{\mu_1\mu_2}\star \F_{\mu_3\mu_4}\star\cdots \star 
\F_{\mu_{2n-1}\mu_{2n}},\cr
b)\quad&{\rm index}\;(-i\Dslash)\,=\,
{(-i)^n\over (4\pi)^n}{1\over n!}\;\int d^{2n}x\, 
\varepsilon^{\mu_1\cdots\mu_{2n}}
\;
{\rm Tr}\,\F_{\mu_1\mu_2}\star \F_{\mu_3\mu_4}\star\cdots \star 
\F_{\mu_{2n-1}\mu_{2n}},\cr
}
$$
respectively, for  cases $a)$ and $b)$. Note the nonlocal character of
the index density in general.

To end this section we will make some comments regarding the structure of
equations~\singletanomaly. The classical 
counterpart of this equation runs
$$
\pr_{\mu}\j^5_{\mu}-\sum_{i,j}( \A_{\mu\,ij}\star
\j^5_{\mu\,ji} - \j^5_{\mu\,ij}\star \A_{\mu\,ji})  = 0.
\eqno\numeq\namelasteq\classical
$$
This is a consequence of the fact that the Lagrangian 
$$
{\cal L}(x)=i\big(\prslash \psi_i\big)_{\alpha}\star\bar{\psi}_{\alpha i}
-i\big(\Aslash_{ij}\star\psi_j\big)_{\alpha}\star\bar{\psi}_{\alpha i} 
\eqno\numeq\namelasteq\lagrangian
$$
is invariant under the following global abelian chiral transformations
$$
\psi'_j\,=\,e^{i\theta\gamma_5}\,\psi_j,\quad
\bar{\psi'}_j\,=\,\bar{\psi}_j\,e^{i\theta\gamma_5};
\eqno\numeq\namelasteq\globaltrans
$$
as can be seen by employing Noether's theorem modified so that the 
noncommutativity of the Moyal product is taken into account. Note that
${\cal L}$ yields (the fermion fields are always Grassmann variables) 
the action in equation~\Uoneaction\ for the group $U(N)$. Now, 
equation~\classical\ can be traded for a simpler equation if instead of
the
current $\j^5_{\mu}$, defined in $a)$ of equation~\singletcurrents , one
uses the current 
$$
\hat{\j}^5_{\mu}\,=\,-\sum_{i}\bar{\psi}_{\beta\,i}
\star\psi_{\alpha\,i}(\gamma_{\mu}\gamma_5)_{\alpha\beta}.
$$
Indeed, taking into account that
$$
\hat{\j}^5_{\mu}\,=\,\j^5_{\mu}\,-\,
[\psi_{\beta\,i},\bar{\psi}_{\alpha\,i}]_{+} 
(\gamma_{\mu}\gamma_5)_{\alpha\beta},
$$
where $[f,g]_{+}=f\star g+ g\star f$, and imposing the equation of 
motion of the fields, one readily shows that
$$
\pr_{\mu}\hat{\j}^5_{\mu}\,=\,
\pr_{\mu}\j^5_{\mu}-\sum_{i,j}( \A_{\mu\,ij}\star
\j^5_{\mu\,ji} - \j^5_{\mu\,ij}\star \A_{\mu\,ji}).
\eqno\numeq\namelasteq\simpler
$$
Hence, equations~\classical\ and \simpler\ lead to 
$$ 
\pr_{\mu}\hat{\j}^5_{\mu}=0,
$$
i.e., the current $\hat{\j}^5_{\mu}$ is classically conserved. 
This conservation can be derived, by using Noether's theorem, 
from the invariance  of the Lagrangian 
$$
\hat{{\cal L}}(x)\,=\,-i\bar{\psi}_{\alpha i}\star\big
(\prslash \psi_i\big)_{\alpha}
+i\bar{\psi}_{\alpha i}\star\big(\Aslash_{ij}\star\psi_j\big)_{\alpha}
\eqno\numeq\namelasteq\hattedlagrangian
$$
under the global transformations defined in equation~\globaltrans .

Notice that both ${\cal L}$ and $\hat{{\cal L}}$, defined in 
equations~\lagrangian\ and ~\hattedlagrangian , respectively,   
yield the same action since
they differ by a Moyal bracket (i.e., a total derivative) 
of the fields and its derivatives:
$$
\hat{{\cal L}}\,=\,{\cal L}+i[\big(\prslash \psi_i\big)_{\alpha}
-\big(\Aslash_{ij}\star\psi_j\big)_{\alpha},\bar{\psi}_{\alpha i}]. 
$$
Putting it all together we conclude that if the equations of motion hold
at the
quantum level the right hand side of $a)$ of equation~\singletanomaly\ can
be recast into the form
$$
\big<\pr_{\mu}\hat{\j}^5_{\mu}\big> =  
2{(-i)^{n-1}\over (4\pi)^n}{1\over n!} \varepsilon^{\mu_1\cdots\mu_{2n}}
\;
{\rm Tr}\,\F_{\mu_1\mu_2}\star \F_{\mu_3\mu_4}\star\cdots \star 
\F_{\mu_{2n-1}\mu_{2n}}.
$$
Similar 
arguments can be put forward for case $b)$ of equation~\singletanomaly.

\section {3. The Nonabelian Anomaly on Noncommutative ${\rm{I\!R}}^{4}$}

In this section we discuss the lack of gauge invariance of the 
effective action, $\Gamma[\A]$,  of Weyl fermions coupled to U(N) gauge
fields on noncommutative  ${\rm {I\!R}}^{4}$. Let $\psi_{R}$ be a
right-handed (say) fermion, i.e., $\psi_{R}=P_{+}\psi$, $\psi$ being a
Euclidean Dirac fermion, and
$P_{+}={1\over 2}(1+\gamma_5)$. In noncommutative ${\rm {I\!R}}^{4}$
there are two basic infinitesimal gauge transformations of $\psi_{R}$
under  the gauge group U(N), namely
$$
a)\;
(\delta_{\theta}\psi_{R})_i=\theta_{ij}\star\psi_{R\,j},\quad 
b)\;(\delta_{\theta}\psi_R)_i=-\psi_{R\,j}\star\theta_{ji},
\eqno\numeq\namelasteq\nonabelianchiraltrans
$$
where $\theta_{ij}=-\theta^{*}_{ji}$ and $i,j= 1,\dots,$N are the $U(N)$
matrix indices. To each of these gauge transformations there is
associated a Dirac operator twisted with U(N) gauge field.  These Dirac
operators act on $\psi_{R}$ as follows
$$
a)\;(\Dslash\, \psi_{R})_i = ( \prslash\psi_{R\,i} - 
\Aslash_{ij}\star\psi_{R\,j}),\; \quad
b)\;(\Dslash\, \psi_{R})_i = ( \prslash\psi_{R\,i} + 
\gamma_{\mu}\psi_{R\,j}\star \A_{\mu\, ji}),
\eqno\numeq\namelasteq\nonabeliandiracoperators
$$
where the gauge field $\A_{\mu}$ is an antihermitian ${\rm N}\times{\rm
N}$ matrix. Each Dirac operator give rise to a classical action which
can be written  in the form
$$
{\rm S}\;=\;\intd^{4}x\; \bar{\psi}\,i\Dslash(\A)_{+}\,\psi,
\eqno\numeq\namelasteq\RHaction
$$ 
where $\Dslash(\A)_{+}=\Dslash(A) P_{+}$. 

Since $\Dslash(\A) P_{+}$ has not a well-defined eigenvalue problem, the 
partition function obtained from $S$ in equation~\RHaction\ cannot be
defined as the  determinant of $\Dslash_{+}$. Hence, we cannot express
$\Gamma[A]$, the  effective action of the right-handed fermion
$\psi_{R}$ in the U(N) background  field
$\A_{\mu}$, in terms of the determinant of $\Dslash(\A)_{+}$. This is an
unwelcomed feature of the action $\RHaction$. In a classic paper 
\cite{\LAGPG} Alvarez-Gaum\'e and Ginsparg taught us how to define 
$\Gamma[A]$ in terms of a determinant. Following these authors we shall
replace $\Dslash(\A)_{+}$ in equation~\RHaction\  with a non-hermitian
operator $\hat{D(\A)}$ ---or simply $\hat{D}$--- defined as follows
$$
a)\;(\hat{D}\, \psi)_i = \prslash\psi_{i} - 
\Aslash_{ij}P_{+}\star\psi_{j},\; \quad
b)\;(\hat{D}\, \psi)_i =\prslash\psi_{i} + 
\gamma_{\mu}P_{+}\psi_{j}\star \A_{\mu\, ji}.
\eqno\numeq\namelasteq\ellipticoperators
$$
Notice that the operators in $a)$ and $b)$ correspond, respectively, to
the  Dirac operator in $a)$ and $b)$ of~\nonabeliandiracoperators. The 
operators in $a)$ and $b)$ of equation~\ellipticoperators\ have a
formally  well-defined eigenvalue problem since they are still elliptic
operators (the  noncommutative character of the space does not modify
the principal symbol of  these operators as regards to their
counterparts in commutative Euclidean  space), with compact inverse,
though they are not hermitian. Taking into account that in
equation~\ellipticoperators\ only the right-handed degrees of freedom
couple to the gauge field, we then define $\Gamma[\A]$ as follows
$$
 e^{-\Gamma[\A]}\;=\;\int d\bar{\psi}d\psi\;
e^{-\intd^{4}x\;\bar{\psi}i\hat{D}(\A)\psi}\;=\;{\rm det}\; i\hat{D}(\A).
\eqno\numeq\namelasteq\effectiveaction
$$
 
We next compute the variation of $\Gamma[\A]$ as defined by 
equation~\effectiveaction\ under gauge transformations; we shall give a
detailed discussion only for $\hat{D}(\A)$ as given by $a)$ in
equation~\ellipticoperators. The results for case $b)$ can be easily
retrieved from the results for case $a)$ by making the appropriate
changes. It can be readily seen that the exponential factor in the
integrand of equation~\effectiveaction\ is invariant under the gauge
transformations 
$$
\delta_{\theta} \A_{\mu\,ij} = 
\partial_{\mu}\theta_{ij}-[\A_{\mu},\theta ]_{ij},\quad
(\delta_{\theta}\psi)_i=\theta_{ij}\star\psi_{R\,j},\quad 
(\delta_{\theta}\bar{\psi})_i=-\bar{\psi}_{L\,j}\star\theta_{ji},
\eqno\numeq\namelasteq\transformations
$$
where $\psi_{L}={1\over 2}(1-\gamma_5)\psi$. Hence, $\Gamma[\A]$ can
fail to be gauge invariant only if the fermionic measure fails to do so.

To compute the change of the fermionic measure, $d\mu=d\bar{\psi}d\psi$
under the chiral gauge transformations in equation~\transformations, we
shall define it as done in reference~\cite{\LAGPG}. One introduces
first right, $\{\varphi_n\}$, and left, $\{\chi^{\dag}_n\}$,
eigenfunctions of $i\hat{D}(\A)$ in $a)$ of equation~\ellipticoperators,
defined as follows
$$
i\hat{D}(\A)\varphi_n=\lambda_n \varphi_n,\quad
(i\hat{D}(\A))^{\dag}\chi_n=\lambda^{*}_n\chi_n,\quad\int
d^4x\;\chi^{\dag}_m(x)
\varphi_m(x)=\delta_{nm}.
$$
Then one expands $\psi$ and $\bar{\psi}$ in terms of these eigenfunctions
$$
\psi(x)=\sum_{n}\,a_n\,\varphi_n (x),\quad
\bar{\psi}(x)=\sum_{n}\,\bar{b}_n\,
\chi^{\dag}_n(x);
$$
so that the fermionic measure reads $d\mu=\prod_n\,d\bar{b}_n da_n$. The 
variation of this measure under the gauge transformations in
equation~\transformations\ is given by
$$
{\cal A}(\theta,\A)=-\sum_n\int d^4x\; \big(\chi^{\dag}_{n i}\star
\theta_{ij} 
\star \gamma_5 
\varphi_{n j}\big)(x).
\eqno\numeq\namelasteq\formalanomaly
$$

We end up with the following formal equation
$$
\delta_{\theta}\Gamma[\A]\;=\;{\cal A}(\theta,\A),
$$
which yields the variation under a gauge transformation of the 
effective action as defined in equation~\effectiveaction. However, the
right hand side of equation~\formalanomaly\ is not well-defined and
needs regularization. We redefine ${\cal A}$ always following
reference~\cite{\LAGPG}
$$
\eqalignno{
{\cal A}(\theta,\A)\,&=\,-\int d^4 x\;\lim_{\Lambda\rightarrow\infty}\;
 \sum_{m}\chi^{\dag}_{n\alpha i}\star
\theta_{ij}\star\gamma_{5\,\alpha\beta} e^{-{\lambda_n^2\over
\Lambda^2}}\varphi_{n \beta j}\cr
\,&=\,-\int d^4 x\;
\lim_{\Lambda\rightarrow\infty}\;
 \sum_{m}\chi^{\dag}_{n\alpha i}\star
\theta_{ij}\star\gamma_{5\,\alpha\beta}
\big(e^{\hat{D}^2\over \Lambda^2}\varphi_{n}\big)_{\beta\, j} \cr
\,&=\,-\int d^4 x\;\lim_{\Lambda\rightarrow\infty}\;
 \sum_{m}\theta_{ij}\star\gamma_{5\,\alpha\beta}\bigg(
\big(e^{\hat{D}^2\over \Lambda^2}\varphi_{n}\big)_{\beta j}
\star\chi^{\dag}_{n\alpha i}\bigg).
&\numeq\cr}
\namelasteq{\defininganomaly}
$$
Expanding $\varphi_n$ and $\chi^{\dag}_n$ in plane waves, we recast 
the expression~\defininganomaly\ into the form
$$
{\cal A}(\theta,\A) \,=\, -\int d^4 x\;\theta_{ij}(x)\; 
\lim_{\Lambda\rightarrow\infty}\;
 \int {d^4 p\over (2\pi)^4}{\rm tr}\,\gamma_{5}\bigg(
\big(e^{\hat{D}^2\over \Lambda^2}e^{ipx}\big)_{ji}\star
e^{-ipx}\big)\bigg).
$$
A little algebra
turns the previous equation into the following one
$$
\eqalign{{\cal A}(\theta,\A)\,=\,-\int d^4 x\;\theta_{ij}(x)\;
\lim_{\Lambda\rightarrow\infty}\;
 \int {d^4 p\over (2\pi)^4}\biggl[&{\rm tr}\,\bigg(P_{+}
\big(e^{-{i\prslash i\Dslash\over \Lambda^2}}e^{ipx}\big)_{ji}\star
e^{-ipx}\big)\bigg)-\cr
&{\rm tr}\,\bigg(P_{-}
\big(e^{- {i\Dslash i\prslash \over \Lambda^2}}e^{ipx}\big)_{ji}\star
e^{-ipx}\big)\bigg)\biggr].\cr
}
$$
Here $i\Dslash$ denotes the Dirac operator introduced in  
equation~\nonabeliandiracoperators\ $a)$. That leads to
$$
\eqalign{{\cal A}(\theta,\A)\,=-\int d^4 x\;\theta_{ij}(x)\;
\lim_{\Lambda\rightarrow\infty}\;
 \int {d^4 p\over (2\pi)^4}\biggl[&{\rm tr}\,\bigg(P_{+}
\big(e^{(i\pslash-\prslash)^2-(i\pslash-\prslash)\Aslash\star\over\Lambda^2
 }{{\rm I}\!\!{\rm I}}\big)_{ji}\bigg)-\cr
&{\rm tr}\,\bigg(P_{-}
\big(e^{(i\pslash-\prslash)^2-\Aslash\star(i\pslash-\prslash)\over\Lambda^2
 }{{\rm I}\!\!{\rm I}}\big)_{ji}\big)\bigg)\biggr],\cr
}
$$
with an obvious notation. A long but straightforward computation
finally yields
$$
{\cal A}(\theta,\A)\,=\,{1\over 24\pi^2}\,{\rm Tr}\int d^4 x\,
\varepsilon_{\mu_1\mu_2\mu_3\mu_4}\,\theta\,\partial_{\mu_1}\bigl[
\A_{\mu_2}\star\partial_{\mu_3}\A_{\mu_4}- {1\over
2}\A_{\mu_2}\star\A_{\mu_3}\star\A_{\mu_4}\bigr]\;+\;\delta_{\theta}
{\cal C}(\Lambda,\A,\theta).
\eqno\numeq\namelasteq\nonabeliananomaly
$$
Here ${\cal C}(\Lambda,\A,\theta)$ stands for an even parity funcional 
which is a polynomial ---in the Moyal product--- of the gauge field and
its 
derivatives, and it is quadraticaly divergent with $\Lambda$.  
This even parity contribution gives rise to a renormalization of
the effective action $\Gamma [A]$. 

The odd parity contribution to ${\cal A}(\theta,\A)$ is the intrinsic
form of the nonabelian anomaly in noncommutative ${\rm {I\!R}}^{4}$. It
cannot
be expressed as the gauge variation of an integrated polynomial in the
Moyal product of the fields and its derivatives as long as the $U(N)$
groups come in direct sums of the fundamental representation. We note
here that this is the case in the Connes-Lott and Chamseddine-Connes
noncommutative geometry formulations of the Standard
Model~\cite{\Cordelia}.
We close this section by remarking that the anomaly we have obtained is
the
so-called consistent anomaly, for it satisfy the Wess-Zumino consistency
conditions~\cite{\WZ}
$$
\delta_{\lambda}{\cal A}(\theta,\A)-\delta_{\theta}{\cal A}(\lambda,\A)=
-{\cal A}([\lambda,\theta],\A),
$$
where $[\lambda,\theta]_{ij}=\lambda_{ik}\star\theta_{kj}-
\theta_{ik}\star\lambda_{kj}$. One can of course pass to the gauge
covariant formulation by adding to the nonabelian chiral current a 
suitable non-covariant polynomial in $\A$. 

The part quadratic in the gauge potentials in equation~\nonabeliananomaly\
can be rewritten as
$$
\eqalign{
&-\,{1\over 96\pi^2}\,\int d^4 x\,
\varepsilon_{\mu_1\mu_2\mu_3\mu_4}\,{\rm Tr}\,T^a\,[T^b,T^c]\,
\partial_{\mu_1}\theta^{a}\bigl[
\A^b_{\mu_2}\star\partial_{\mu_3}\A^c_{\mu_4}-
\partial_{\mu_3}\A^c_{\mu_4}\star\A^b_{\mu_2} \bigr]\cr
&-\,{1\over 96\pi^2}\,\int d^4 x\,
\varepsilon_{\mu_1\mu_2\mu_3\mu_4}\,{\rm Tr}\,T^a\,\{T^b,T^c\}\,
\partial_{\mu_1}\theta^{a}\bigl[
\A^b_{\mu_2}\star\partial_{\mu_3}\A^c_{\mu_4}+
\partial_{\mu_3}\A^c_{\mu_4}\star\A^b_{\mu_2} \bigr].\cr
}
$$
The first term is a new contribution that does not vanish in the noncommutative
case.

\section {4. Conclusions}

The form of the nonabelian anomaly in equation~\nonabeliananomaly\ is
a na\"{\i}ve $\star$-deformation of the consistent anomaly in
commutative Euclidean space. However, this apparently inocuous operation
has physical consequences. Indeed, the vanishing of the anomalous
contribution from triangle diagrams in commutative  euclidean space
demands the famous anomaly cancellation condition ${\rm 
Tr}\,T^a\{T^b,T^c\}=0$ to hold. The reader is invited to check  
that the term left after imposing ${\rm Tr}\,T^a\{T^b,T^c\}=0$ in the anomalous gauge
field quadratic part of equation~\nonabeliananomaly\ is not (modulo
$\A^3$) a $\delta_{\theta}$-exact polynomial in the Moyal product of the
gauge field and its derivatives. Therefore, in noncommutative ${\rm
{I\!R}}^{4}$ the anomalous contribution from the triangle diagrams
(i.e., the contribution in~\nonabeliananomaly\ which is quadratic in the
gauge fields) will vanish if, and only if, ${\rm  Tr}\,T^a T^b T^c =0$.
This is a consequence of the
noncommutative character of the Moyal product. 

As a further curiosity, notice that the pure  SU(2) contribution to the
nonabelian anomaly, which vanishes on a commutative ${\rm {I\!R}}^{4}$,
does not vanish for a noncommutative ${\rm {I\!R}}^{4}$.  Recall that
the left-handed fermions of the Standard Model are $SU(2)$ doublets.

We conclude that if our theory on noncommutative ${\rm{I\!R}}^{4}$ has a
right-handed fermion transforming under the $U(N)$ group as in $a)$ of
equation~\nonabelianchiraltrans, the only chance of having an anomaly
free theory is to have a left-handed fermion transforming in the same
manner. We see below that having a right-handed fermion transforming as
in $b)$ of equation~\nonabelianchiraltrans\ can be turned into having a
left-handed fermion transforming as in $a)$ of
equation~\nonabelianchiraltrans\ by means of charge
conjugation~\cite{\Sheikh}. The theory will thus be a vector theory. 
This result has been already established for the group $U(1)$
in~\cite{\Hayakawa}.

Therefore, to close we shall compute the behaviour under a gauge 
transformation  of the effective action $\Gamma[A]$ when $i\hat{D}$ in
equation~\effectiveaction\ is given by the elliptic operator in $b)$ of 
equation~\ellipticoperators. This situation corresponds to having a
right-handed fermion transforming as given in
$b)$ equation~\nonabelianchiraltrans. One realizes that the situation at
hand can be converted into the previous one by expressing the former in
terms of
$\star_{-\omega}$ by using $f\star_{\omega} g = g \star_{-\omega} f$  and
performing the substitutions $\A_{\mu\,ij}\rightarrow - \A_{\mu\,ji}$.
Thus, we can use formula~\nonabeliananomaly\ to obtain 
$$
\delta_{\theta}\Gamma[A]\,=\,
\,-{1\over 24\pi^2}\,{\rm Tr}\int d^4 x\,
\varepsilon_{\mu_1\mu_2\mu_3\mu_4}\,\theta\,\partial_{\mu_1}\bigl[
\A_{\mu_2}\star\partial_{\mu_3}\A_{\mu_4}-
{1\over 2}\A_{\mu_2}\star\A_{\mu_3}\star\A_{\mu_4}\bigr]
+\;\delta_{\theta}
{\cal C}(\Lambda,\A,\theta).
$$
Notice that the intrinsic anomaly given by the previous equation
and the intrinsic anomaly given in equation~\nonabeliananomaly\ have
oposite sign. Hence, a theory with two right-handed fermions
transforming, respectively,  as defined by
$a)$ and $b)$ in equation~\nonabelianchiraltrans\ will be anomaly free.
This
theory is equivalent to a theory with a right-handed fermion and a
left-handed fermion both transforming as given by $a)$ in
equation~\nonabelianchiraltrans: just define
$\psi_{R}=\tilde{\psi}^c_{L}$, where
$c$ stands for charge conjugation, for a right-handed fermion
$\tilde{\psi}$ transforming as in $a)$ of
equation~\nonabelianchiraltrans.
\bigskip
\section{Acknowledgments}

This work has been partially supported by CICyT under grant PB98-0842. 
The work of J.M. Gracia-Bond\'{\i}a has also been partially supported by
UCM's ``Sab\'aticos Complutense 1999''.

\bigskip\bigskip

\section{References}

\frenchspacing

\refno\ABJ.
S. Adler, Phys. Rev. 177 (1969) 2426. J. Bell and R. Jackiw,
Nuovo Cimento A 60 (1969) 47. W. Bardeen, Phys. Rev. 184 (1969)
1848.

\refno\Marshak.
S.B. Treiman, R. Jackiw, B. Zumino and E. Witten, {\it Current Algebra and 
Anomalies}, World Scientific, Singapore, 1985.
R. E. Marshak, {\it Conceptual Foundations of Modern
Particle Physics}, World Scientific, Singapore, 1993. 

\refno\Booss.
B. Booss and D.D. Bleecker, {\it Topology and Analysis: The Atiyah-Singer
Index
Formula and Gauge-Theoretic Physics}, Springer-Verlag, 1985.

\refno\Filk.
T. Filk, Phys. Lett. B376 (1996) 53.
M. Berkooz, Phys. Lett. B430 (1998) 237.
J. C. V\'arilly and J. M. Gracia-Bond\'{\i}a, Int. J. Mod.
Phys. A14 (1999) 1305;  {\tt hep-th/9804001}.
M. Chaichian, A. Demichev and P. Presnajder, 
``Quantum field theory on noncommutative space-time and
the persistence of ultraviolet divergences'',{\tt hep-th
9904132}. C. P. Mart\'{\i}n and D. S\'anchez Ruiz, Phys.
Rev. Lett. 83 (1999) 476; {\tt hep-th/9903187}.
M.M. Sheikh-Jabbari, JHEP 9906 (1999) 015.
T. Krajewski and R. Wulkenhaar, ``Perturbative Quantum Gauge Fields on the
noncommutative Torus'', {\tt hep-th 9903187}.

\refno\Minwallaetal.
S. Minwalla, M. Van Raamsdonk and N. Seiberg,
``Noncommutative perturbative dynamics", {\tt hep-th/9912172}.
M. Hayakawa ``Perturbative analysis on infrared aspects of 
noncommutative QED on ${\rm {I\!R}}^{4}$'', {\tt hep-th/9912094}.
Y. Aref'eva, D. M. Belov and A. S. Koshelev, ``Two-loop diagrams in 
noncommutative $\phi^4_4$ theory'', {\tt hep-th 9912075},
``A note on UV/IR for noncommutative complex scalar field'',
 {\tt hep-th/0001215}.
Harald Grosse, Thomas Krajewski, Raimar Wulkenhaar, 
``Renormalization of noncommutative gauge Yang-Mills theories: a simple
  example'', {\tt Hep-th 0001182}.
A. Matusis, L. Susskind and N. Toumbas,   
``The IR/UV connection in the noncommutative gauge theories'', 
{\tt hep-th/0002075}.

\refno\miscellanea.
M. Alishahiha, Y. Oz, M. M. Sheikh-Jabbari, JHEP 9911:007, 1999.
A. Armoni, ``A note on noncommutative orbifold field theories'',
{\tt hep-th/9910031}.
J. Ambjorn, Y. M. Makeenko, J. Nishimura, R. J. Szabo, JHEP 9911 (1999)
029.
I. Chepelev and  R. Roiban, ``Renormalization of Quantum field Theories on
noncommutative ${\rm {I\!R}}^{D}$. 1. Scalars. {\tt hep-th 9911098}.
H. B. Benaom, ``Perturbative BF Yang-Mills Theory on noncommutative  
${\rm {I\!R}}^{4}$'', {\tt hep-th/9912036}.
G. Arcioni and M. A. Vazquez-Mozo, JHEP 0001 (2000) 028.
W. Fischler, E. Gorbatov, A. Kashani-Poor, S. Paban, P. Pouliot and J.
Gomis,
``Evidence for winding states in noncommutative quantum field theory'', 
{\tt hep-th/0002067}.
J. Madore, S. Schraml, P. Schupp and J. Wess, ``Gauge theory on
noncommutative
spaces'', {\tt hep-th 0001203}.
 
\refno\stringt.
A. Connes, M. R. Douglas and A. Schwarz, JHEP 9802 (1998) 003.
N. Nekrasov and A. Schwarz, Commun. Math. Phys. 198 (1998) 689.
C. Hofman and E. Verlinde, Nucl. Phys. B547 (1999) 157.
J. M. Maldacena and J. G. Russo, JHEP 9909 (1999) 025.
M. Li and Y-S. Wu, ``Holography and Noncommutative Yang-Mills'', 
{\tt hep-th/9909085}.
J. L. F. Barbon and  E. Rabinovici, JHEP 9912 (1999) 017.

\refno\SW. 
N. Seiberg and E. Witten, JHEP 9909  (1999) 032.

\refno\ArdalanS.
F. Ardalan and N. Sadooghi, ``Axial anomaly in
noncommutative QED on ${\rm {I\!R}}^{4}$", {\tt hep-th/0002143}.

\refno\Fujikawa.
K. Fujikawa, Phys. Rev. D 21 10 (1980) 2848.

\refno\Manolo.
M. Asorey, ``Introduction to gauge anomalies", DFTUZ preprint, 1985.

\refno\Phobos.
J. M. Gracia-Bond\'{\i}a and J. C. V\'arilly, J. Math. Phys. 29 (1988)
869.

\refno\LAGPG. L. Alvarez-Gaum\'e and P. Ginsparg, Nucl. Phys. B 243
(1984) 449.

\refno\Cordelia.
C. P. Mart\'{\i}n, J. M. Gracia-Bond\'{\i}a and J. C. V\'arilly, 
Phys. Rep. 294 (1998) 363.

\refno\WZ. J. Wess and B. Zumino, Phys. Lett. B 37 (1971) 95. 

\refno\Sheikh. M.M. Sheikh-Jabbari, ``Discrete Symmetries (C, P, T) in
Noncommutative Field  Theories'', {\tt hep-th/0001167}.

\refno\Hayakawa. M. Hayakawa, ``Perturbative analysis on infrared and
ultraviolet aspects of  noncommutative QED on ${\rm {I\!R}}^{4}$'', {\tt
hep-th/9912167}.

\end